\documentclass[lettersize,journal]{IEEEtran}
\usepackage{amsmath,amsfonts}
\usepackage{algorithm}
\usepackage{array}
\usepackage[caption=false,font=normalsize,labelfont=sf,textfont=sf]{subfig}
\usepackage{textcomp}
\usepackage{stfloats}
\usepackage{url}
\usepackage{verbatim}
\usepackage{graphicx}
\usepackage{cite}
\usepackage{cuted}
\usepackage{amsmath,amssymb,amsfonts}
\usepackage{graphicx}
\usepackage{textcomp}
\usepackage{soul}
\usepackage{setspace}
\usepackage{amsthm}
\usepackage{todonotes}
\usepackage{menukeys}
\usepackage{soul}
\usepackage{bbm}

\usepackage[compact]{titlesec}
\titlespacing{\section}{0pt}{2ex}{1ex}
\titlespacing{\subsection}{0pt}{2ex}{1ex}
\usepackage[table]{xcolor}
\definecolor{darkgreen}{rgb}{0,0.6,0}
\usepackage{tabularx}
\usepackage{nccmath}
\usepackage[strict]{changepage}

\usepackage{cite}
\usepackage[colorlinks=true, allcolors=blue]{hyperref}
\usepackage{tikz}
\usepackage{setspace}
\usetikzlibrary{arrows,shapes}
\usepackage{nicematrix}
\usepackage{arydshln}

\usepackage{algpseudocode}

\makeatletter
\newcommand{\manuallabel}[2]{\phantomsection\def\@currentlabel{#2}\label{#1}}
\makeatother

\begin{document}

\title{Autonomous and Distributed Synchronization and Restoration of an Islanded Network of Microgrids}

\author{Ahmed Saad Al-Karsani,~\IEEEmembership{Student Member,~IEEE,} Maryam Khanbaghi,~\IEEEmembership{Senior Member,~IEEE}
\thanks{Ahmed Saad Al-Karsani and Maryam Khanbaghi are with Santa Clara University, 500 El Camino Real, Santa Clara, CA, USA. (email: \{ahsaad, mkhanbaghi\}@scu.edu)}
}



\maketitle

\begin{abstract}
The transition towards clean energy and the introduction of Inverter-Based Resources (IBRs) are leading to the formation of Microgrids (MGs) and Networks of MGs (NMGs). MGs and NMGs can operate autonomously in islanded mode, which requires Grid-Forming (GFM) IBRs that can perform black start, synchronization, restoration and regulation. However, such IBRs can face synchronization instability issues, which might be worsened by inadequate secondary level frequency and voltage regulation. Accordingly, we propose an autonomous and distributed synchronization and restoration scheme using Distributed-Averaging Proportional-Integral (DAPI) control. To validate the proposed method, we model and simulate a high-fidelity islanded and modified IEEE 123 bus system, modeled as an NMG consisting of 7 MGs. The MATLAB\textsuperscript{\textregistered}/Simulink\textsuperscript{\textregistered} simulation results demonstrate an effective autonomous soft-start, synchronization, connection and regulation procedure using DAPI control and distributed breaker operation logic. 
\end{abstract}

\begin{IEEEkeywords}
Network of Microgrids (NMGs), Grid-Forming (GFM) Inverter-Based Resources (IBRs), DAPI
\end{IEEEkeywords}

\section{Introduction}
The introduction of Renewable Energy Sources (RES) is transforming the structure of the power grid. RES can be used in bulk form or as Distributed Energy Resources (DERs), changing the grid from its centralized to a distributed structure. Although this adds complexity in terms of balancing between supply and demand, it introduces enhanced reliability and resilience due to the possibility of forming Microgrids (MGs). MGs can operate in either grid-connected or islanded mode, giving them an edge in terms of quality of service. For further flexibility, reliability and resilience, one school of thought proposes the concept of Networks of MGs (NMGs) \cite{b1}.

The operation of NMGs in islanded mode faces many challenges. RES are typically interfaced with the grid through power-electronic devices, i.e., Inverter-Based Resources (IBRs). Such devices differ in operation principle from conventional Synchronous Machines (SMs), as they synthesize an AC signal with a specific voltage and frequency based on measurements and setpoints. This means that there is no rotational inertia, making it more sensitive to disturbances and more susceptible to failure. Therefore, studies are ongoing on the design and implementation of effective control schemes for IBRs, particularly within the scope of hierarchical control. Hierarchical control involves stabilization and power sharing at the primary level, frequency and voltage regulation at the secondary level and energy management at the tertiary level.

In terms of the NMG's frequency and voltage, IBRs can operate in either Grid-Forming (GFM) or Grid-Following (GFL) modes. Islanded MGs and NMGs require GFM sources in order to operate autonomously. The most common ones are droop-controlled IBRs at the primary level, establishing a linear relation between active power and frequency as well as reactive power and voltage, respectively. Additionally, frequency and voltage regulation might be necessary to avoid excessive deviations and/or tripping events. In the literature, various distributed secondary level control schemes have been proposed using consensus control, most notably Distributed-Averaging Proportional-Integral (DAPI) control \cite{b2}. The scheme ensures distributed frequency and voltage regulation with power sharing for droop-controlled IBRs without the need for supervisory control, and avoids having any single points of failure. 

During extreme scenarios such as blackouts, GFM IBRs are expected to have black start capability, per the UNIFI consortium guidelines \cite{b3}. Specifically, the IBRs must accommodate for safe black start, restoration and synchronization processes. Pan \textit{et al.}'s \cite{b4} review of IBR-based black start lists active power and frequency control as well as reactive power and voltage control as potential challenges during the process. Naderi \textit{et al.}'s \cite{b5} study on NMGs synchronization stability has shown that inrush power can occur due to small voltage, frequency and phase differences at the Point of Common Coupling (PCC). Additionally, they found that secondary level control plays an essential role in synchronization stability. 

Other works address MGs synchronization with bulk power grids using secondary level (consensus) control, such as that of Sun \textit{et al.} \cite{b6}, Du \textit{et al.} \cite{b7} and Tomás-Martín \textit{et al.} \cite{b8}. In terms of our area of interest for islanded NMGs, Shah \textit{et al.} \cite{b9} presented an MG consensus-based synchronization scheme at multiple points of interconnection, with the option of leaderless consensus control when there is no bulk grid connection. However, reactive power sharing and voltage regulation were not addressed. Naderi \textit{et al.}'s \cite{b5} work concerning NMGs with distributed (consensus) control has extensively shown that secondary level control is essential for synchronization stability between MGs and/or NMGs. Additionally, their findings have shown that synchronization constraints in IEEE Standard 1547 may not be strict enough to avoid instability in weak grids such as islanded NMGs. 

A largely unexplored area worth mentioning is that during islanded MG and NMG operation, the existence of distributed secondary level control means that there is a possibility of autonomous synchronization and connection operation. Upreti and Mohammed \cite{b10} have shown that automatic synchronization schemes can be safely and economically implemented in MGs using standard relays and communications. Their scheme, termed the Autosynchronization (A25A) algorithm, is currently in use in practice. Nonetheless, secondary level regulation and power sharing are yet to be explored, which might be necessary regarding weak grids such as islanded NMGs at the distribution level.

Another adjacent challenge is the realistic large-scale modeling of such islanded NMGs at the distribution level. While Alam \textit{et al.} \cite{b11} developed a benchmark NMG test system, however, the studies citing the work have generally used it for static simulation (power flow) studies. Pombo \textit{et al.} \cite{b12} modeled and simulated a benchmark system based on the two islands of Cape Verde, but only using transfer function equivalent models for generation sources. Suresh \textit{et al.} \cite{b13,b14} developed a detailed model of the IEEE 123 bus system in MATLAB\textsuperscript{\textregistered}/Simulink\textsuperscript{\textregistered}. The system is an unbalanced and multi-phase system, fed from the upstream of the substation. They altered the IEEE 123 bus system to include solar and storage grid-connected IBRs. Banerjee \textit{et al.} \cite{b15} and Seo \textit{et al.} \cite{b16} address GFM IBR black start using realistic simulations, but secondary control is yet to be addressed.

In summary, there is a need for GFM IBRs with black start services that can also achieve autonomous and distributed synchronization and regulation in islanded NMGs. Additionally, accurate high-fidelity modeling of NMGs is also necessary for realistic simulation and validation. Accordingly, this paper proposes the following contributions:  
\begin{enumerate}
    \item Implement a distributed and leaderless phase synchronization scheme for each IBR, with autonomous synchronization and connection process within the NMG.
    \item Perform soft-start, synchronization and connection while maintaining frequency and voltage regulation as well as power sharing within the NMG.
    \item Validate and simulate the proposed scheme using a large-scale NMG consisting of 7 MGs and IBRs, based on the unbalanced IEEE 123 bus test feeder system containing GFM IBRs with droop, DAPI and lower level control. 
\end{enumerate}

The rest of this paper is organized as follows: Section \ref{Section II} introduces the IBR modeling and control scheme, followed by the black start and synchronization scheme in Section \ref{Section III}. The simulation results are shown in Section \ref{Section IV}, before listing our concluding remarks and future work in Section \ref{Section V}.

\section{Inverter Modeling and Control}\label{Section II}
In this section, we introduce an IBR model consisting of a droop-controlled inverter with DAPI frequency and voltage control, aggregated into an islanded NMG system. The IBR model is illustrated in Fig. \ref{Fig. 1}, comprising an Battery Energy Storage System (BESS), an inverter, and an LCL filter. Lower level outer voltage and inner current loops and Pulse Width Modulation (PWM) are utilized, and due to page constraints, we refer to \cite{b2,b5} for details. The primary and secondary level control schemes are explained in the following subsections.

\begin{figure}\centering\manuallabel{Fig. 1}{1}
  \includegraphics[width=\columnwidth]{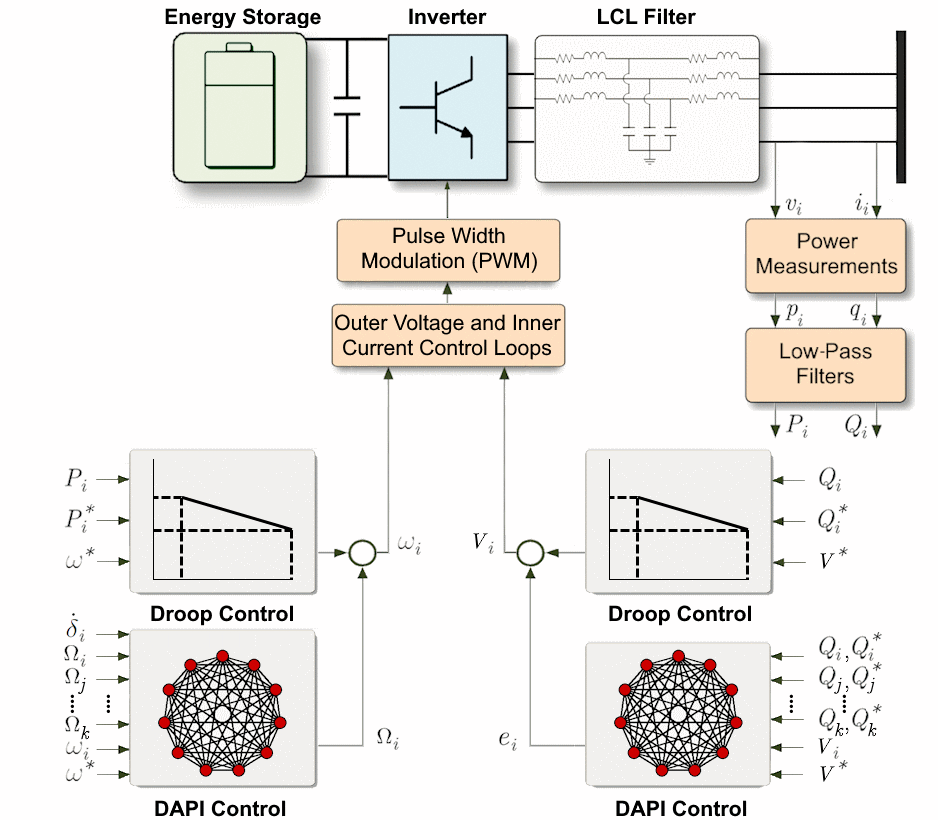}
  \caption{The GFM IBR interfaced with the BESS, where each local DAPI control receives $j^{th}$ to $k^{th}$ DAPI variables from neighboring IBRs.}
  \vspace{-15pt}
\end{figure}

  
\subsection{Primary Level Control}
We assume the existence of droop-controlled GFM IBRs interfaced with BESS. The droop-controlled inverter model is governed by the following equations \cite{b6}:
\begin{equation}\label{(1)}\dot{\delta}_i(t) = \omega_i(t) - \omega^*\end{equation}
\begin{equation}\label{(2)}\omega_i(t) = \omega^*  -m_i(P_i(t)-P_i^*)\end{equation}
\begin{equation}\label{(3)}V_i(t) = V^*  -n_i(Q_i(t)-Q_i^*)\end{equation}
where $\delta_i(t)$ is the phase angle, $\omega_i(t)$ and $\omega^*$ are the actual and setpoint frequency, $m_i$ is the active power-frequency droop gain and $P_i(t)$ and $P_i^*$ are the actual and reference active power output. Also, $V_i(t)$ and $V^*$ are the actual and setpoint voltage, $n_i$ is the reactive power-voltage droop gain and $Q_i(t)$ and $Q_i^*$ are the actual and reference reactive power output.

Since droop control is a form of proportional control, it results in steady-state errors that must be compensated for using added secondary control terms as follows:
\begin{equation}\label{(4)}\omega_i(t) = \omega^*  -m_i(P_i(t)-P_i^*) + \Omega_i(t)\end{equation}
\begin{equation}\label{(5)}V_i(t) = V^*  -n_i(Q_i(t)-Q_i^*) + e_i(t)\end{equation}
where $\Omega_i(t)$ and $e_i(t)$ are the $i^{th}$ DAPI frequency and voltage consensus variables, explained in the following subsection.

\subsection{Secondary Level Control}
For frequency and voltage regulation, we employ the DAPI control scheme from \cite{b2}. This assumes a fixed NMG communication topology modeled as a graph \cite{b2,b7}. In this scheme, regulation is achieved along with power sharing for droop-controlled IBRs using the following consensus dynamics \cite{b2}:
\begin{equation}\label{(6)} k_{i,i}\dot{\Omega}_i(t) = -\eta_i\Delta \omega_i(t) - \sum_{j=1}^Na_{ij}(\Omega_i(t) - \Omega_j(t))\end{equation}
\begin{equation}\label{(7)}\kappa_{i,i}\dot{e}_i(t) = -\xi_i\Delta V_i(t) - \sum_{j=1}^Nb_{ij}\left(\frac{Q_i(t)}{Q_i^*} - \frac{Q_j(t)}{Q_j^*}\right)\end{equation}
where $\Delta \omega_i = \omega_i - \omega^*$ and $\Delta V_i = V_i - V^*$, $k_{i,i}$ and $\kappa_{i,i}$ are the DAPI frequency and voltage inverse integral gains, $a_{ij}$ and $b_{ij}$ are the respective communication gains (1 if there's a link and 0 otherwise), $\eta_i,\xi_i$ are the frequency and voltage deviation gains and $N$ is the total number of NMG IBRs. DAPI control achieves frequency regulation and active power sharing, with a tradeoff between voltage regulation and reactive power sharing based on NMG loads, lines and topology characteristics.

During black start when each MG is islanded, $a_{ij}$ and $b_{ij}$ are set as zero, achieving decentralized frequency and voltage regulation. However, this may not be enough for synchronization, as there might be considerable phase offsets between IBRs. To address that, we introduce phase consensus in the following subsection.

\subsection{Phase Consensus Synchronization}
For initial phase synchronization in islanded NMGs assuming sufficiently low output impedance, we propose the following altered dynamics for each IBR's phase $\delta_i$ in \eqref{(1)}:\vspace{-1pt}
\begin{equation}\label{(8)} \dot{\delta}_i(t) = \Delta \omega_i(t) - \sum_{j=1}^Nd_{ij}(\theta_{i}(t) - \theta_{j}(t))\end{equation}
where $\theta_i$ and $\theta_j$ are local and neighboring MG PCC angles and $d_{ij}$ is the respective communication gain. This follows similar consensus dynamics as \eqref{(6)} for phase synchronization at sufficiently low MG PCC output impedance. By initially adding $\dot{\delta}_i$ to \eqref{(6)}, it contributes towards distributed synchronization in MGs and NMGs that are islanded from the bulk power grid. We note that \eqref{(8)} relies on the assumption that there is a sufficient communication link, i.e., a connected graph with a direct or indirect link between every IBR in the NMG.

Following the modeling of the GFM IBR with primary level droop control and secondary level DAPI control, in the next section, we will introduce the IBR black start and synchronization requirements along with the proposed scheme.


\section{Inverter Black Start and Synchronization}\label{Section III}
Following a blackout, a black start is required to energize the buses and restore power delivery. In the case of MGs and NMGs, this requires a GFM source, i.e., droop-controlled IBRs, to safely handle the cold start and load pickup processes. The UNIFI specifications for GFM IBRs outlined the requirements for black start capability, namely \cite{b3}:
\begin{enumerate}
    \item Internally start and establish the voltage.
    \item Safely supply the system within IBR limits, potentially using voltage ramping.
    \item Provide a steady voltage reference for smooth restoration and synchronization.
    \item Optionally achieve collective black start, if operating in parallel with other GFM IBRs.
\end{enumerate}

In this work, we cover the aforementioned requirements, with the caveat of performing black start for each individual MG before synchronization and connection across the NMG. This is due to the assumption that each IBR can adequately supply its local MG load, per the UNIFI specifications \cite{b3}. 

Additionally, per Pan \textit{et al.}'s review, we aim to address the challenges of active power and frequency control, as well as reactive power and voltage control during such restoration process \cite{b4}. We also note Naderi \textit{et al.}'s \cite{b5} findings showing that even small nonzero voltage, frequency and phase differences can lead to small-signal instability. Furthermore, they conclude that secondary level control plays an important role in synchronization stability. With that in mind, we consider the following additional criteria:
\begin{enumerate}
    \item Secondary level frequency and voltage regulation before synchronization and connection.
    \item Active power sharing and frequency regulation after synchronization and connection.
    \item Reactive power sharing, voltage regulation and the tradeoff between them after synchronization and connection.
\end{enumerate}

For synchronization criteria, Table \ref{Table 1} shows the synchronization parameter limits from the IEEE 1547-2018 standards. Per \cite{b5}, instability can occur even for criteria within the IEEE 1547 standards. In our work, we set the tolerances for $|\Delta \omega|$, $|\Delta V|$, and $|\Delta \theta|$ to be less than the aforementioned standard limits. In this case, a low tolerance can also help ensure a steady state is reached before the breaker closing process. It is worth noting that for the proposed method, $\Delta \theta$ refers to the PCC phase consensus term in \eqref{(8)}.

To automate the synchronization and connection process, we propose the logic outlined in Fig. \ref{Fig. 2}, Fig. \ref{Fig. 3} and Fig. \ref{Fig. 4}. Fig. \ref{Fig. 2} shows each IBR's local synchronization condition check $IBR_{SYNC,LOCAL,i}$ regarding frequency, voltage and phase differences. An AND gate is used to output 1 if the check is passed. Following that process, a second stage AND gate in Fig. \ref{Fig. 3} is used for the IBR to determine if both its local and neighboring synchronization checks have passed. Neighboring IBRs are decided by MGs' physical connections. So, for the breaker to close, we require the IBRs in the interconnecting MGs to exchange a synchronization confirmation check, before the breaker connecting those MGs receives the okay through its relay. The second stage check is used as a redundancy to ensure simultaneous breaker closings across the NMG. Those second-stage outputs $IBR_{SYNC,i}$ can be passed to the breaker relay adjacent to each MG (and by extension IBR), shown in Fig. \ref{Fig. 4}. If the adjacent IBRs' second-stage outputs are 1, then the SR latch sets the output to the breaker relay as 1 ($BREAKER_{CLOSE}$).

\begin{table}[!b]\manuallabel{Table 1}{I}\vspace{-12pt}
\caption{Synchronization Parameter Limits}
\centering
\begin{tabular}{c|c|c|c}
\hline
\hline
\multicolumn{4}{c}{\textit{IEEE 1547-2018 Standards}} \\
\hline
\textit{DER Rating (kVA)} & \textit{$|\Delta f|$} & \textit{$|\Delta V|$} & \textit{$|\Delta \theta|$} \\
\hline
$< 500$ & $0.3$ Hz & $10\%$ & $20^\circ$ \\
$500-1500$ & $0.2$ Hz & $5\%$ & $15^\circ$ \\
$>1500$ & $0.1$ Hz & $3\%$ & $10^\circ$ \\
\hline
\hline
\multicolumn{4}{c}{\textit{Proposed Work}} \\
\hline
\textit{DER Rating (kVA)} & \textit{$|\Delta f|$} & \textit{$|\Delta V|$} & \textit{$\sum |\Delta \theta|$} \\
\hline
$-$ & $0.01$ Hz & $1\%$ & $2.5^\circ$ \\
\hline
\hline
\end{tabular}
\label{tab1}
\end{table}

\begin{figure}[t!]\vspace{-4pt}
  \centering
  \begin{minipage}[t]{0.48\textwidth}
    \centering
    \includegraphics[width=\textwidth]{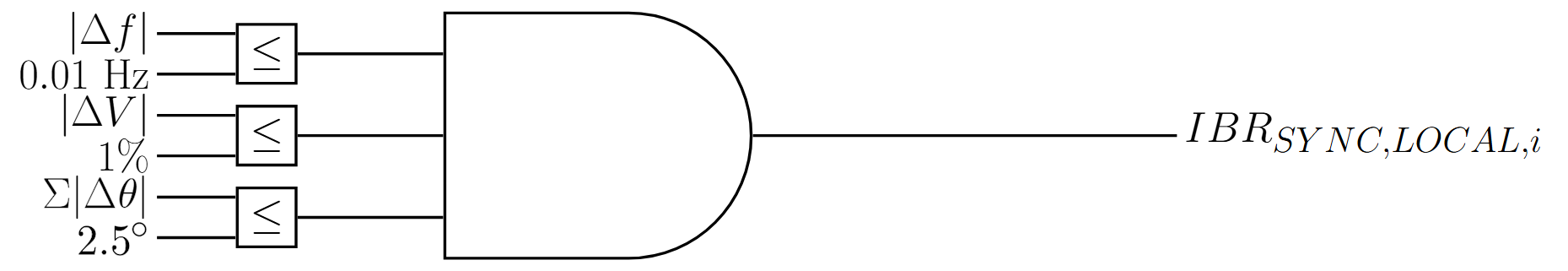}
    \caption{%
      Each $i^{th}$ IBR's local synchronization conditions check.
    }
    \manuallabel{Fig. 2}{2}
  \end{minipage}%

  \vspace{8pt} 

  \begin{minipage}[t]{0.48\textwidth}
    \centering
    \includegraphics[width=\textwidth]{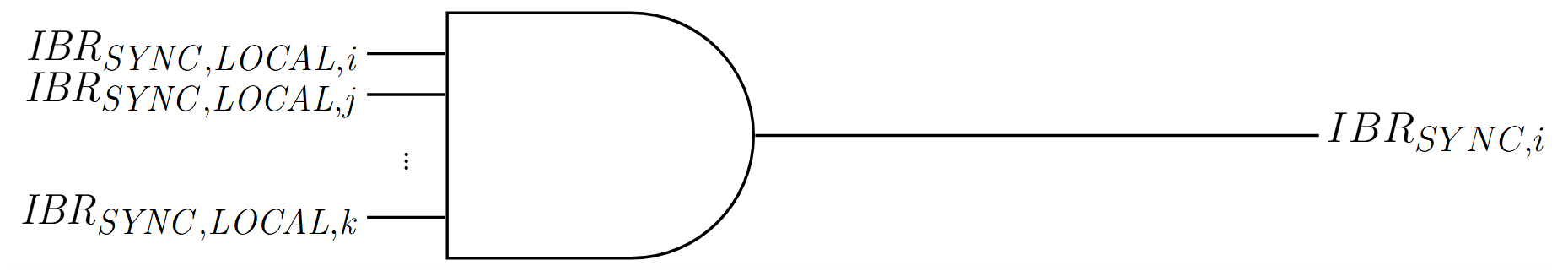} 
    \caption{%
      Each $i^{th}$ IBR's local and $j^{th},\dots,k^{th}$ neighboring IBRs/MGs synchronization conditions check.
    }
    \manuallabel{Fig. 3}{3}
  \end{minipage}%

    \vspace{8pt} 

  \begin{minipage}[t]{0.48\textwidth}
    \centering
    \includegraphics[width=\textwidth]{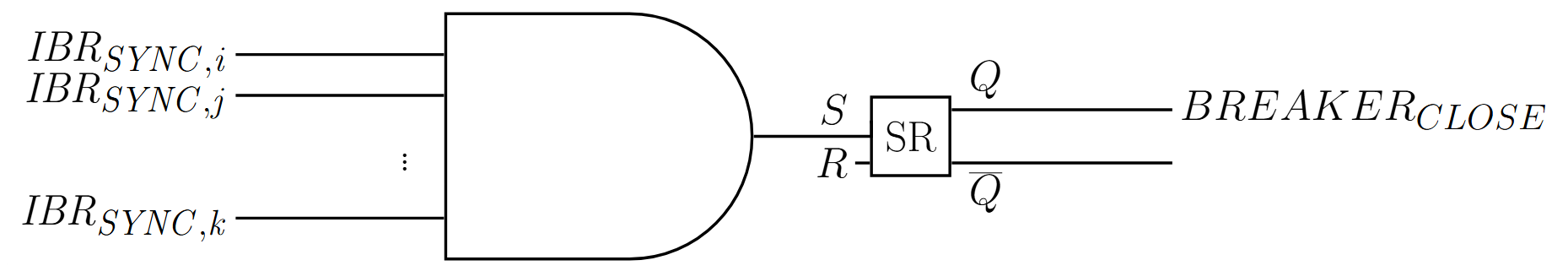} 
    \caption{%
      Breaker connection logic based on adjacent $i^{th},j^{th},\dots,k^{th}$ IBR/MG synchronization checks.
    }
    \manuallabel{Fig. 4}{4}
  \end{minipage}%
      \vspace{-14pt}
\end{figure}

In the following section, we validate the proposed scheme using simulations in MATLAB\textsuperscript{\textregistered}/Simulink\textsuperscript{\textregistered}.

\section{Simulation Results}\label{Section IV}
In this section, we validate the proposed synchronization and restoration scheme using secondary level DAPI control, distributed phase synchronization and distributed synchronization check logic. An NMG is modeled based on an islanded IEEE 123-bus feeder system in MATLAB\textsuperscript{\textregistered}/Simulink\textsuperscript{\textregistered} (SimPowerSystems library per \cite{b13,b14}), shown in Fig. \ref{Fig. 5}. Each blue dashed line shows a communication link between IBRs.

The NMG topology is based on the proposed design of Zhou \textit{et al.} \cite{b17}, where the system is isolated from the upstream feeder. The system consists of 7 MGs. Each MG has a single GFM IBR, as described in Section \ref{Section II}. Table \ref{Table 2} shows the relevant NMG parameters. Each IBR is modeled in detail with an IGBT/diode universal bridge block, along with primary level, secondary level and lower level outer voltage and inner current control schemes. Nonetheless, it is acknowledged that average-based modeling of inverters is typically sufficient for primary level and secondary level control studies.

\subsection{Simulation Results}
Fig. \ref{Fig. 6} shows the simulation results in terms of active power, frequency, reactive power, voltage and phase respectively. At $t = 0$ sec, we simulate a simplified black start operation, where each MG is islanded and each IBR performs a soft-start process ramping $V^*$ from $0$ V to $480\frac{\sqrt{2}}{\sqrt{3}}$ V within $0.5$ seconds. Decentralized frequency and voltage regulation is performed at the secondary level, to regulate the frequency and voltage around their setpoint values $f^*$ and $V^*$. Additionally, the power setpoint values $P^*$ and $Q^*$ are fixed to be the respective islanded MG served load. Due to the parameter limits defined in Table \ref{Table 1}, phase and voltage regulation occur within a couple of seconds, and frequency synchronization occurs last.

At $t = 3$ sec, each IBR's local synchronization conditions signal $IBR_{SYNC,LOCAL,i}$, and consequently every $IBR_{SYNC,i}$ signal value becomes 1, triggering breaker closures across the NMG. $P^*$ and $Q^*$ become the respective average of the sum of every islanded IBR's setpoint, which can be determined by the operator at the tertiary level beforehand, or shared and computed among IBRs using their rated apparent power $S_{rated,i}$ and participation factor $\alpha_i = \frac{S_{rated,i}}{\sum_{i=1}^NS_{rated,i}}$. Following transients for approximately 1 second, each IBR outputs an equal amount of active power while regulating the frequency to $60$ Hz. Also, for $\xi = 0.05$, there is a tradeoff between voltage regulation (within $\pm5\%$ of $V^*$ at steady state) and reactive power regulation. We note that without DAPI control, reactive power circulation occurs following the NMG connection process.

\begin{figure}[!t]\manuallabel{Fig. 5}{5}\vspace{-6pt}
    \begin{center}
  \includegraphics[width=0.9\columnwidth]{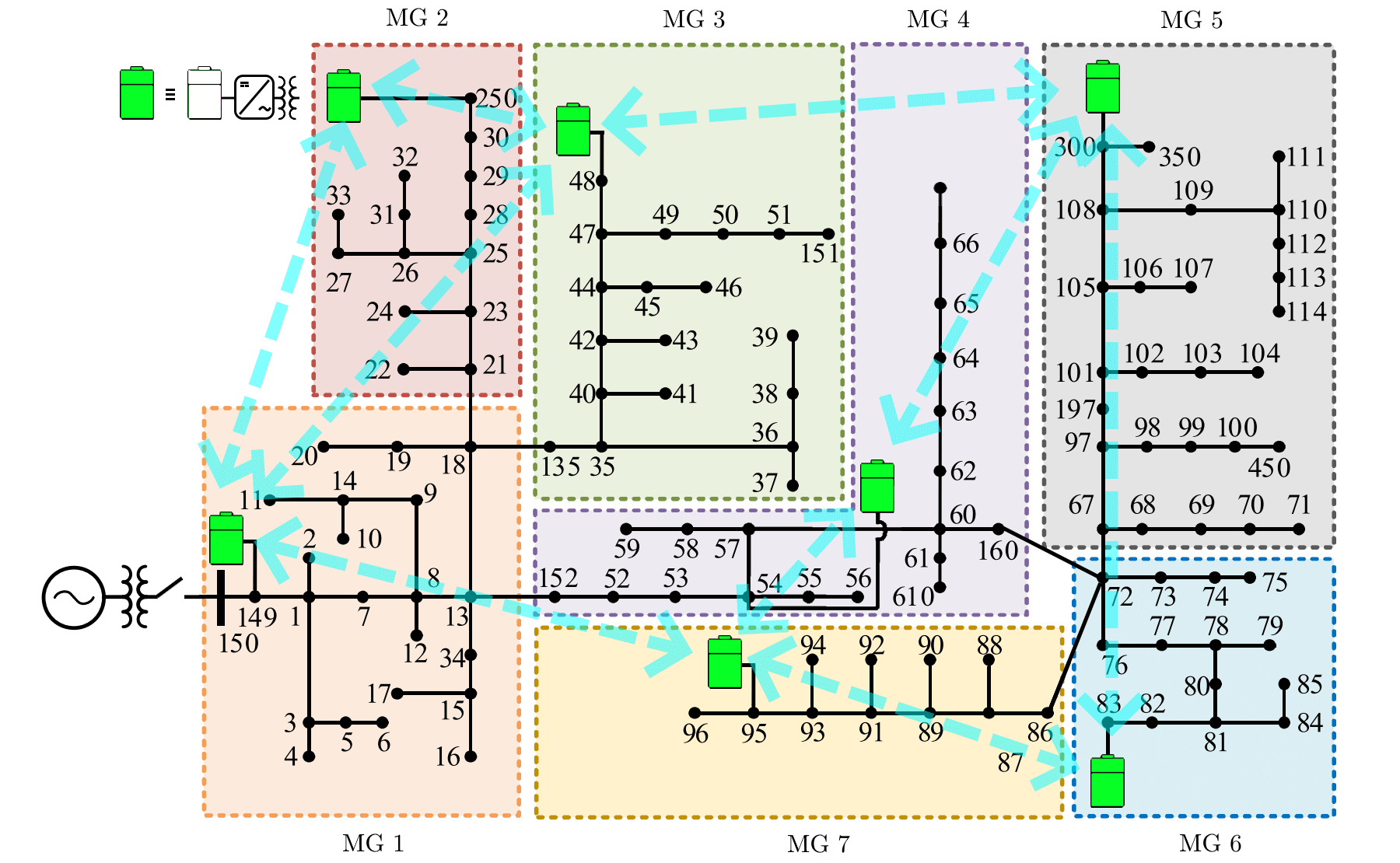}
  \caption{The NMG model based on a modified unbalanced IEEE 123 bus system design per \cite{b17}. The blue dashed lines indicate communication links.}
    \end{center}
    \vspace{-5pt}
\end{figure}

\begin{table}\manuallabel{Table 2}{II}\vspace{-8pt}
\caption{NMG Relevant Parameters}
\centering
\begin{tabular}{c c c c}
\hline
\hline
\textit{Parameter} & \textit{Symbol} & \textit{Value} \\
\hline
IBR Setpoint Frequency & $f^*$ & $60$ Hz \\
\hline
IBR Setpoint Voltage & $V^*$ & $480\frac{\sqrt{2}}{\sqrt{3}}$ V \\
\hline
Active Power-Frequency Droop & $m$ & $\frac{1}{1000000}$ $\frac{rad/s}{W}$ \\
\hline
Reactive Power-Voltage Droop & $n$ & $\frac{48\frac{\sqrt{2}}{\sqrt{3}}}{1000000}$ $\frac{V}{VAr}$ \\
\hline
Communication Link Time Step & $T_{com}$ & $1$ sec \\
\hline
Communication Link Gain & $a,b,d$ & $0$ or $1$ \\
\hline
Voltage Deviation Gain & $\xi$ & $0.05$ \\
\hline
\hline
\end{tabular}
\vspace{-10pt}
\label{tab1}
\end{table}

\begin{figure*}[h!]\centering
  \includegraphics[width=\textwidth]{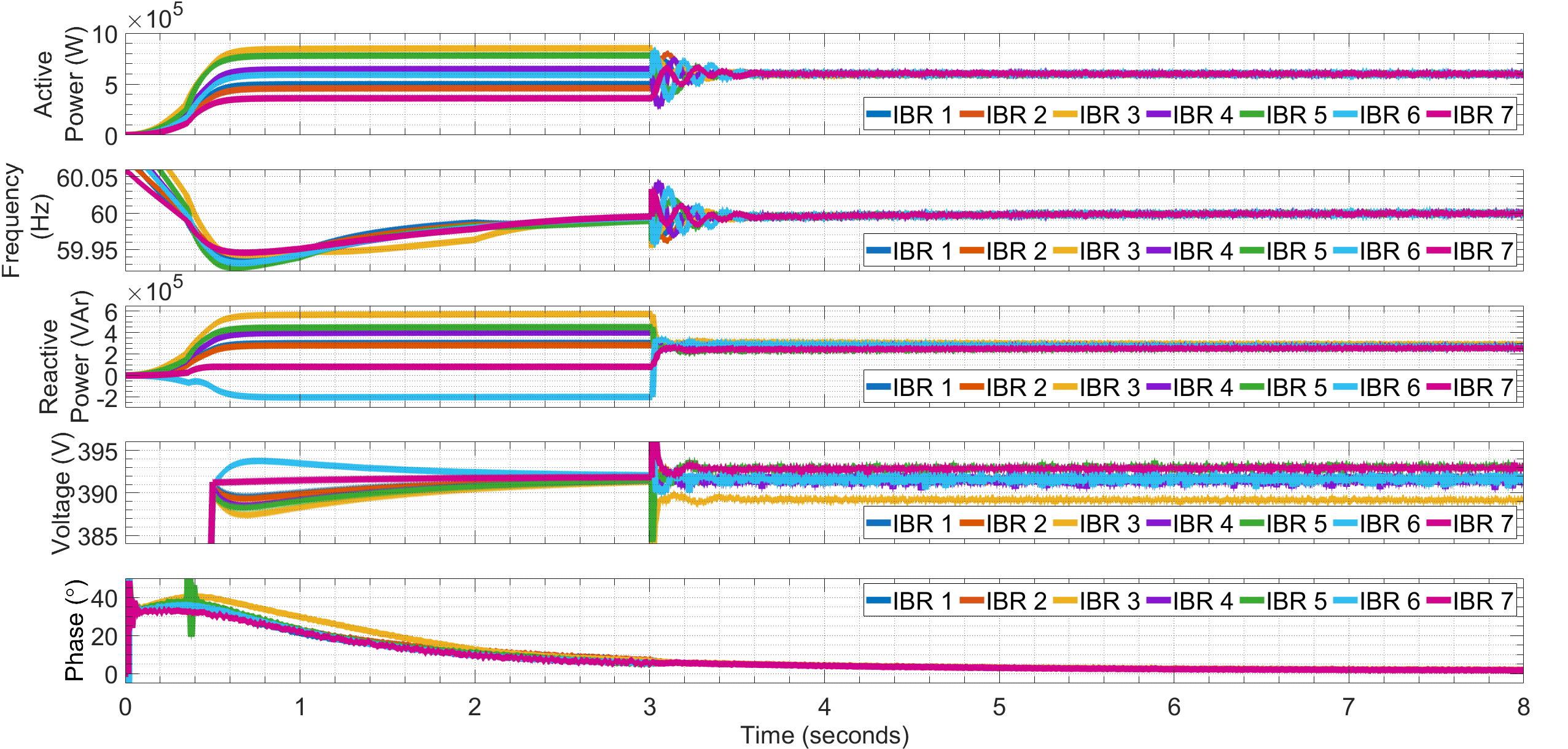}
  \caption{\label{Fig. 6}From top to bottom: IBR active power, frequency, reactive power, voltage and phase from 0 to 10 seconds.}
  \vspace{-5pt}
\end{figure*}

\subsection{Results Discussion}
The simulation shows promising results in terms of autonomous and distributed synchronization and restoration of the NMG. Nonetheless, prior simulations have shown that instability can occur following closures if the primary and/or secondary level control time step is too large. In the illustrated results, an 11.1 ms time step was chosen for primary level and secondary level control (plus a 1-second time step for receiving neighboring $\Omega_j$ and $Q_j$ parameters). In the case of operating primary level and secondary level control at 22.2 ms, the system becomes unstable following the breakers' closure. Robust control schemes can be deployed to deal with longer time steps and time delays.

We also remark that the communication link time delay of 1 second is for a time scale resembling conventional secondary level control, i.e., Automatic Generation Control (AGC). However, recent advances in low-latency communication protocols, i.e., IEC 61850 Generic Object Oriented Substation Event (GOOSE), might open the possibility for distributed secondary level control closer to the primary level control range (set to be 11.1 milliseconds in this study). This protocol is already used within MG protection schemes in practice (such as S\&C Ameren MG \cite{b18}). Generally, GOOSE messaging can be used for communication between breaker relays and IBRs.

Lastly, the unbalanced loads within the IEEE 123 bus system worsen the voltage and current unbalance factors and result in distortions, especially after NMG connection. This is outside of the scope of this study, and left for future work.

\section{Conclusions and Future Work}\label{Section V}
An autonomous and distributed synchronization and restoration scheme has been proposed for islanded NMGs. The scheme employs DAPI control with distributed phase consensus for phase synchronization. Additionally, an automated and distributed connection scheme has been proposed for the reconnection of islanded MGs, based on local and neighboring synchronization conditions checks. The scheme has been validated using a high-fidelity large-scale model of an NMG (based on the IEEE 123 bus test feeder system and consisting of 7 MGs), and simulated in MATLAB\textsuperscript{\textregistered}/Simulink\textsuperscript{\textregistered} (using SimPowerSystems library). Future work involves testing during faults and communication failures, improved voltage and power delivery during unbalanced conditions and robust control, especially under longer time steps and/or potential time delays within primary level and secondary level control.

\end{document}